\documentclass[reprint,noshowpacs,noshowkeys,prd,balancelastpage,onecolumn,nofootinbib]{revtex4}
\usepackage{amsfonts}
\usepackage{mdframed}
\usepackage{amssymb}
\usepackage{footnote}
\usepackage{amsmath}
\usepackage{graphicx}
\usepackage{float}
\usepackage[font={footnotesize,it}]{caption}
\usepackage[utf8]{inputenc}
\usepackage{natbib}
\usepackage{tikz}
\usepackage{tikz-3dplot}
\usepackage[colorlinks=true,
            linkcolor=blue,
            urlcolor=blue,
            citecolor=red]{hyperref}

\begin{document}

\title{Colliding waves in a model of nonlinear electrodynamics}
\author{S. Habib Mazharimousavi}
\email{habib.mazhari@emu.edu.tr}
\author{M. Halilsoy}
\email{mustafa.halilsoy@emu.edu.tr}
\affiliation{Department of Physics, Faculty of Arts and Sciences, Eastern Mediterranean
University, Famagusta, North Cyprus via Mersin 10, Turkey}
\date{\today }

\begin{abstract}
Bell-Szekeres (BS) solution for colliding electromagnetic waves in
Einstein-Maxwell (EM) theory describes also colliding waves in nonlinear
electrodynamics (NED) with an emergent cosmological constant. Our NED model
covers the first leading orders to the well-known Heisenberg-Euler (HE) type
in a particular gauge of pure magnetic field. Prior to the problem of
collision we obtain dyonic solution for the considered NED theory in a
conformally flat spacetime which has both electric and magnetic fields with
constant invariants. Our sole finding is that null currents inevitably arise
in the process of collision of plane waves in the HE type NED theory.
\end{abstract}

\maketitle

\section{Introduction}

During 1970s and 1980s collision of plane waves was a popular subject to
explore the nonlinear features of interaction in general relativity (GR) 
\cite{JBG}. The collision process ended mostly in spacetime singularities
and rarely with unstable horizons. It is known that the four experimental
tests of GR are all verified through weak field approximations whereas the
strong field effects, such as collision of waves remains an effect to be
seen. In modified theories and their higher order nonlinear corrections
finding exact solutions become challenging enough to push the topic out of
fashion. In this study we consider collision of electromagnetic (em) waves
in nonlinear electrodynamics (NED). Our model of NED is described by the
Heisenberg-Euler (HE) Lagrangian $L\left( P,Q\right) $ in its leading order
terms where $P=\frac{1}{4}F_{\mu \nu }F^{\mu \nu }$ and $Q=\frac{1}{4}F_{\mu
\nu }\tilde{F}^{\mu \nu }$ are the Maxwell invariants. The HE effective
action, was first derived by W. Heisenberg and H. Euler \cite{EH} for
constant electromagnetic fields at one-loop order in the fluctuating quantum
fields. It provides quantum corrections to the linear Maxwell's theory in
terms of nonlinear interactions. Experimentally, for a high intensity field,
the HE is a more accurate classical approximation of quantum electrodynamic
than linear Maxwell's theory \cite{IZ,9}. We note that our notation and
equations follow from Ref. \cite{Oliveria} (see also \cite{NED1}).

The action of our model is given by ($G=1$)%
\begin{equation}
I=-\frac{1}{4\pi }\int \sqrt{-g}\left( \frac{R}{4\pi }+2\Lambda +L\left(
P,Q\right) \right) d^{4}x  \label{1}
\end{equation}%
where $R$ is the Ricci scalar, $\Lambda $ is the cosmological constant and%
\begin{equation}
L\left( P,Q\right) =P+\beta P^{2}+\gamma Q^{2}  \label{2}
\end{equation}%
in which the positive constants $\gamma $ $\ $and $\beta $ are related by $%
\gamma =\frac{7}{4}\beta .$ Such a representation may be considered as a
post-Maxwellian approach to the HE theory \cite{Denisov}. Let us add that,
in a general nonlinear electrodynamics (2) is at the lowest order described
by a Lorentz invariant parity conserving Lagrangian correction. We must also
add that the action (1) covers more general classes of theories but since
our interest in this study is restricted by conformal flatness this will set 
$R=0.$ The nonlinear Maxwell (NLM) equation reads%
\begin{equation}
d\left( L_{P}\mathbf{\tilde{F}}+L_{Q}\mathbf{F}\right) =0  \label{3}
\end{equation}%
in which the field $2$-form is chosen to be 
\begin{equation}
\mathbf{F}=E\left( r\right) dt\wedge dr+B_{0}\sin \theta d\theta \wedge
d\phi ,  \label{4}
\end{equation}%
with its dual $\mathbf{\tilde{F},}$ where $B_{0}$ is a magnetic constant and 
$E\left( r\right) $ is to be determined. In the sequel our metric ansatz
will be conformally flat so that expectedly the scalar curvature will be
zero. The gravitational field equations are%
\begin{equation}
G_{\mu }^{\nu }+\Lambda \delta _{\mu }^{\nu }=8\pi T_{\mu }^{\nu }  \label{5}
\end{equation}%
where $G_{\mu }^{\nu }$ is the Einstein tensor and the energy momentum
tensor $T_{\mu }^{\nu }$ of NED is given by%
\begin{equation}
T_{\mu }^{\nu }=-\frac{1}{4\pi }\left( L\delta _{\mu }^{\nu }-L_{P}F_{\mu
\lambda }F^{\nu \lambda }-QL_{Q}\delta _{\mu }^{\nu }\right)   \label{6}
\end{equation}%
in which $L_{P}=\frac{\partial L}{\partial P}$ and $L_{Q}=\frac{\partial L}{%
\partial Q}$. In Section II we shall find the general spherically symmetric
integral of the theory with a cosmological constant such that both
invariants $P$ and $Q$ become constant. The constancy of $P$ and $Q$ is the
main cause that we can easily introduce cosmological constant to the
problem. In Section III we make a particular choice of gauge to simplify our
ansatz as much as possible before we transform our metric into the space of
double null coordinates. This will provide us the case with $P\neq 0$ and $%
Q=0$ which is the simplest choice as far as the leading terms of the HE
theory is concerned.

\section{An integral of the theory with conformally flat ansatz}

Our conformally flat, static spacetime is described by the line element 
\begin{equation}
ds^{2}=\Phi ^{2}\left( -dt^{2}+dr^{2}+r^{2}d\theta ^{2}+r^{2}\sin ^{2}\theta
d\phi ^{2}\right) ,  \label{7}
\end{equation}%
where $\Phi =\Phi \left( r\right) $ is the only unknown metric function to
be found. The energy momentum tensor (\ref{6}) implies%
\begin{equation}
T_{t}^{t}=T_{r}^{r}=-\frac{1}{4\pi }\left( L-L_{P}F_{tr}F^{tr}-QL_{Q}\right) 
\label{8}
\end{equation}%
and 
\begin{equation}
T_{\theta }^{\theta }=T_{\phi }^{\phi }=-\frac{1}{4\pi }\left(
L-L_{P}F_{\theta \phi }F^{\theta \phi }-QL_{Q}\right)   \label{9}
\end{equation}%
which upon (\ref{5}) and (\ref{8}) one finds%
\begin{equation}
G_{t}^{t}=G_{r}^{r}  \label{10}
\end{equation}%
and%
\begin{equation}
G_{\theta }^{\theta }=G_{\phi }^{\phi }.  \label{11}
\end{equation}%
The explicit form of the Einstein tensor's components are given by%
\begin{equation}
G_{t}^{t}=\frac{1}{\Phi ^{4}r}\left( 2r\Phi \Phi ^{\prime \prime }+4\Phi
\Phi ^{\prime }-r\Phi ^{\prime 2}\right) ,  \label{tt}
\end{equation}%
\begin{equation}
G_{r}^{r}=\frac{\Phi ^{\prime }}{\Phi ^{4}r}\left( 4\Phi +3r\Phi ^{\prime
}\right)   \label{rr}
\end{equation}%
and%
\begin{equation}
G_{\theta }^{\theta }=G_{\phi }^{\phi }=\frac{1}{\Phi ^{4}r}\left( 2r\Phi
\Phi ^{\prime \prime }+2\Phi \Phi ^{\prime }-r\Phi ^{\prime 2}\right) .
\label{thth}
\end{equation}%
The constraint (\ref{10}) on $G_{t}^{t}$ and $G_{r}^{r}$ implies%
\begin{equation}
\Phi \Phi ^{\prime \prime }=2\Phi ^{\prime 2}  \label{eq}
\end{equation}%
which amounts to an exact solution for the metric function obtained as%
\begin{equation}
\Phi =\frac{\alpha }{r+r_{0}}  \label{12}
\end{equation}%
in which $\alpha $ and $r_{0}$ are two integration constants. Let's comment
that in (\ref{12}) we set $r_{0}=0$ so that the line element (\ref{7})
becomes the Bertotti-Robinson (BR) spacetime \cite{BR}. On the other hand
from $\mathbf{F}$ given in (\ref{4}) the dual of $\mathbf{F}$ follows as%
\begin{equation}
\mathbf{\tilde{F}}=-\frac{B_{0}}{r^{2}}dt\wedge dr+E\left( r\right)
r^{2}\sin \theta d\theta \wedge d\phi   \label{13}
\end{equation}%
so that one obtains 
\begin{equation}
P=\frac{1}{2\Phi ^{4}}\left( \frac{B_{0}^{2}}{r^{4}}-E^{2}\right) 
\label{14}
\end{equation}%
and%
\begin{equation}
Q=\frac{EB_{0}}{r^{2}\Phi ^{4}}.  \label{15}
\end{equation}%
The NLM equation (\ref{3}), then, yields%
\begin{equation}
L_{P}Er^{2}+L_{Q}B_{0}=q  \label{16}
\end{equation}%
in which $L_{P}=1+2\beta P$ and $L_{Q}=2\gamma Q$ and $q$ is an integration
constant. With the help of (\ref{12}) (with $r_{0}=0$), (\ref{14}) and (\ref%
{15}), the latter equation admits a solution for the electric field given by%
\begin{equation}
E=\frac{E_{0}}{r^{2}}  \label{17}
\end{equation}%
in which $E_{0}$ is a constant satisfying 
\begin{equation}
\left( 1+\frac{B_{0}^{2}}{\alpha ^{4}}\left( \beta +2\gamma \right) \right)
E_{0}-\frac{\beta }{\alpha ^{4}}E_{0}^{3}=q.  \label{18}
\end{equation}%
Furthermore, with the exact solution for the metric function $\Phi \left(
r\right) =\frac{\alpha }{r}$, from (\ref{tt}), (\ref{rr}) and (\ref{thth})
we obtain the Einstein's tensor given by%
\begin{equation}
G_{\mu }^{\nu }=\frac{1}{\alpha ^{2}}diag\left[ -1,-1,1,1\right]   \label{19}
\end{equation}%
which upon considering the $tt$ and $\theta \theta $ components of the
Einstein's field equations (\ref{5}) we obtain%
\begin{equation}
\Lambda =\frac{1}{\alpha ^{2}}+\frac{2\gamma E_{0}^{2}B_{0}^{2}}{\alpha ^{8}}%
-\frac{B_{0}^{2}+E_{0}^{2}}{\alpha ^{4}}+\beta \frac{\left(
E_{0}^{2}-B_{0}^{2}\right) \left( B_{0}^{2}+3E_{0}^{2}\right) }{2\alpha ^{8}}
\label{20}
\end{equation}%
and 
\begin{equation}
\alpha ^{6}-\alpha ^{4}\left( E_{0}^{2}+B_{0}^{2}\right) +\beta \left(
E_{0}^{4}-B_{0}^{4}\right) =0  \label{21}
\end{equation}%
respectively. We note that, the latter two equations are algebraic equations
and solving them simultaneously gives the integration constants $\alpha $
and $q$ (via $E_{0}$ and $B_{0}$) in terms of our theory constants $\beta $
and $\Lambda .$ This completes the spherically symmetric integral of the
action in BR spacetime such that we obtain $P=const.$ and $Q=const.$ from
Eqs. (\ref{14}) and (\ref{15}), respectively. To complete this section, we
study the energy conditions by considering the energy-momentum tensor given
by $T_{\mu }^{\nu }=diag[-\rho ,p_{r},p_{\perp },p_{\perp }]$ in which $\rho
=-T_{t}^{t}$ is the energy density, $p_{r}=T_{r}^{r}$ is the radial pressure
and $p_{\perp }=T_{\theta }^{\theta }=T_{\phi }^{\phi }$ is the lateral
pressure. It is straightforward to show that conservation of energy-momentum
is held, i.e., $\triangledown _{\mu }T^{\mu \nu }=0.$ The null energy
condition (NEC) implies that $\rho +p_{r}\geq 0$ and $\rho +p_{\perp }\geq 0.
$ From (\ref{8}) and (\ref{9}) by imposing (\ref{21}) we find $\rho +p_{r}=0$
and $\rho +p_{\perp }=\frac{1}{4\pi \alpha ^{2}},$ implying that NEC is
satisfied. The weak energy condition (WEC), however, requires $\rho \geq 0$
in addition to NEC. The explicit expression for the energy density is found
to be%
\begin{equation}
\rho =\frac{1}{4\pi }\left( \frac{B_{0}^{2}+E_{0}^{2}}{2\alpha ^{4}}+\frac{%
\beta \left( B_{0}^{2}-E_{0}^{2}\right) \left( B_{0}^{2}+3E_{0}^{2}\right)
-\gamma B_{0}^{2}E_{0}^{2}}{4\alpha ^{8}}\right)   \label{ED1}
\end{equation}%
which upon eliminating $\beta $ (and consequently $\gamma =\frac{7}{4}\beta $%
) from (\ref{21}) when $B_{0}^{2}\neq E_{0}^{2}$ it becomes 
\begin{equation}
\rho =\frac{1}{16\pi \left( \xi -1\right) \alpha ^{2}}\left( \frac{\left(
\xi ^{2}+5\xi +1\right) E_{0}^{2}}{\alpha ^{2}}+\frac{\xi ^{2}-5\xi -3}{\xi
+1}\right)   \label{ED2}
\end{equation}%
where $\xi =\frac{B_{0}^{2}}{E_{0}^{2}}\neq 1.$ In general, from (\ref{ED2})
we observe that $\rho \geq 0$ and $\rho <0$ are both possible implying the
WEC is satisfied for fine tuned parameters.

For the specific cases when $\xi =\frac{B_{0}^{2}}{E_{0}^{2}}=1$ one finds%
\begin{equation}
\alpha ^{2}=2E_{0}^{2}  \label{ED3}
\end{equation}%
and 
\begin{equation}
\rho =\frac{4E_{0}^{2}-\gamma }{64\pi E_{0}^{4}}.  \label{ED4}
\end{equation}%
In the Lagrangian (\ref{2}) $\beta $ and $\gamma $ are positive parameters,
implying that the latter energy density is definitely non-negative if $%
E_{0}^{2}=B_{0}^{2}\leq \frac{\gamma }{4}$ and consequently the WEC is
satisfied.

Furthermore, for the extreme case when $E_{0}=0$ one finds%
\begin{equation}
\rho =\frac{B_{0}^{2}+\alpha ^{2}}{16\pi \alpha ^{4}}  \label{ED5}
\end{equation}%
which is clearly positive. However, in order to satisfy the condition $\beta
=\frac{\alpha ^{4}\left( \alpha ^{2}-B_{0}^{2}\right) }{B_{0}^{4}}>0$, one
finds $\alpha ^{2}>B_{0}^{2}.$ This is of our interest for the next section.
Concerning this particular case, one finds%
\begin{equation}
p_{\perp }=\frac{3\alpha ^{2}-B_{0}^{2}}{16\pi \alpha ^{4}}  \label{ED6}
\end{equation}%
which is positive due to the positive definite of $\beta $. Hence, not only
NEC and WEC are satisfied but also the strong energy condition (SEC) - which
implies the NEC together with $\rho +p_{r}+2p_{\perp }\geq 0$ - is satisfied.

Setting $E_{0}=0,$ in the next section we shall transform this particular
spherical geometry to the double-null coordinates and interpret the solution
obtained as colliding em waves.

\section{The space of colliding waves}

The solution obtained in the previous section is too general with both
constants $E_{0}\neq 0\neq B_{0}$. For this reason we shall make a
particular choice with $E_{0}=0$, so that the entire energy-momentum will be
expressed in terms of the magnetic field $B_{0}.$ In this section, we choose
the magnetic gauge such that the field $2$-form is defined by (from the
potential $1$-form $\mathbf{A}=-B_{0}\cos \theta d\phi $)%
\begin{equation}
F=B_{0}\sin \theta d\theta \wedge d\phi  \label{22}
\end{equation}%
and its dual 
\begin{equation}
\mathbf{\tilde{F}}=-\frac{B_{0}}{r^{2}}dt\wedge dr.  \label{23}
\end{equation}%
With this reduction, the invariants are $P=\frac{B_{0}^{2}}{2\alpha ^{4}}%
=const.$ and $Q=0$ so that the NLM equation which takes the form%
\begin{equation}
d\left( L_{P}\mathbf{\tilde{F}}\right) =0  \label{24}
\end{equation}%
is automatically satisfied. The line element obtained in the previous
section\ (with the choice $\alpha =1$) namely, the BR 
\begin{equation}
ds^{2}=\frac{1}{r^{2}}\left( -dt^{2}+dr^{2}\right) +d\theta ^{2}+\sin
^{2}\theta d\phi ^{2}  \label{25}
\end{equation}%
and the field $2$-form are transformed now by the transformation \cite{JBG}%
\begin{equation}
t+r=\coth \left( \frac{z-y}{2}\right) ,  \label{26}
\end{equation}%
\begin{equation}
t-r=-\tanh \left( \frac{z+y}{2}\right) ,  \label{27}
\end{equation}%
\begin{equation}
\theta =u-v+\frac{\pi }{2}  \label{28}
\end{equation}%
and 
\begin{equation}
\phi =x  \label{29}
\end{equation}%
where $\frac{1}{\cosh z}=\cos \left( u+v\right) .$ Upon this transformation
we obtain the metric (\ref{25}) in double-null coordinates form as 
\begin{equation}
ds^{2}=-4dudv+\cos ^{2}\left( u-v\right) dx^{2}+\cos ^{2}\left( u+v\right)
dy^{2}  \label{30}
\end{equation}%
with the em potential $1$-form, given by 
\begin{equation}
\mathbf{A}=B_{0}\sin \left( u-v\right) dx.  \label{31}
\end{equation}%
We make the shifts now, by replacing $u\rightarrow u\theta \left( u\right) ,$
$v\rightarrow v\theta \left( v\right) $ where $\theta \left( u\right) $, $%
\theta \left( v\right) $ are Heaviside unit step functions, which give 
\begin{equation}
\mathbf{F}=B_{0}\cos \left( u\theta \left( u\right) -v\theta \left( v\right)
\right) \left( \theta \left( u\right) du-\theta \left( v\right) dv\right)
\wedge dx  \label{32}
\end{equation}%
and its dual%
\begin{equation}
\mathbf{\tilde{F}}=B_{0}\cos \left( u\theta \left( u\right) +v\theta \left(
v\right) \right) \left( \theta \left( u\right) du+\theta \left( v\right)
dv\right) \wedge dy.  \label{33}
\end{equation}%
It is observed by virtue of the step functions that for $u<0,v<0$ the Region
I is flat \cite{JBG}. The incoming regions, Region II (for $u>0,v<0$) and
Region III (for $u<0,v>0$) have null em fields both linearly polarized in
the $x$-direction perpendicular to the ($u,v$) plane. As a result in the
interaction region (Region IV) the em waves are also polarized in the $x$%
-direction. By this choice for polarizations we eliminate any birefringence
that will arise in the interaction region. For the interaction region, $u>0,$
$v>0$ (Region IV) we have the non-zero energy-momentum components%
\begin{equation}
T_{uu}=T_{vv}=\frac{B_{0}^{2}}{4\pi }\left( 1+\beta B_{0}^{2}\right)
\label{34}
\end{equation}%
\begin{equation}
T_{uv}=-\beta \frac{B_{0}^{4}}{8\pi },  \label{35}
\end{equation}%
and%
\begin{equation}
T_{xx}=\frac{B_{0}^{2}}{8\pi }\left( 1+\frac{3}{2}\beta B_{0}^{2}\right)
\cos ^{2}\left( u-v\right) ,T_{yy}=-\frac{B_{0}^{2}}{8\pi }\left( 1+\frac{1}{%
2}\beta B_{0}^{2}\right) \cos ^{2}\left( u+v\right)  \label{36}
\end{equation}%
and the emergent cosmological constant 
\begin{equation}
\Lambda =\beta \frac{B_{0}^{4}}{2}.  \label{37}
\end{equation}%
The non-zero Einstein's tensor components for (\ref{30}) are 
\begin{equation}
G_{uu}=G_{vv}=2,  \label{uu}
\end{equation}%
\begin{equation}
G_{xx}=\cos ^{2}\left( u-v\right)  \label{xx}
\end{equation}%
and%
\begin{equation}
G_{yy}=-\cos ^{2}\left( u+v\right) .  \label{yy}
\end{equation}%
These are all obtained from Einstein's equations (\ref{5}) with the
cosmological constant adapted to the metric (\ref{30}). Let us note that,
for the consistency of the field equations we must have the constraint%
\begin{equation}
\beta =\frac{1}{B_{0}^{2}}\left( \frac{1}{B_{0}^{2}}-1\right)  \label{38}
\end{equation}%
which can easily be seen from the equality of right and left sides of the
Einstein equations. Hence, from (\ref{38}) and (\ref{37}) one obtains%
\begin{equation}
\Lambda =\frac{1}{2}\left( 1-B_{0}^{2}\right) .  \label{CC}
\end{equation}%
For strong magnetic field where $B_{0}^{2}\gg 1$ (\ref{CC}) amounts to $%
\Lambda \sim -B_{0}^{2}$ while $\beta \sim -\frac{1}{B_{0}^{2}}.$ Using the
conversion factor to transform the geometric units into SI units one finds%
\begin{equation}
\Lambda \simeq -\frac{4\pi \epsilon _{0}G}{c^{4}}B_{0}^{2}  \label{CC2}
\end{equation}%
in which $G,$ $\epsilon _{0}$ and $c$ are the gravitational constant, vacuum
electric permittivity and speed of light respectively. Considering the
typical value of the cosmological constant i.e., $10^{-52}\frac{1}{m^{2}}$
we estimate the magnitude of $B_{0}$ to be of order $\sim 10T.$ The NLM
equation (\ref{3}) as adapted to the new spacetime now gives 
\begin{equation}
d\left( \mathbf{\tilde{F}}L_{P}\right) =\mathbf{\tilde{J}}\neq 0  \label{39}
\end{equation}%
with the current $3$-form 
\begin{equation}
\mathbf{\tilde{J}}=\beta B_{0}^{3}\left( \theta \left( u\right) \delta
\left( v\right) \cos u-\theta \left( v\right) \delta \left( u\right) \cos
v\right) du\wedge dv\wedge dy  \label{40}
\end{equation}%
in which $\delta \left( u\right) $ and $\delta \left( v\right) $ are Dirac
delta functions which vanish only for $u>0$ and $v>0$. It is seen that this
null current vanishes only for $\beta =0$ (for $B_{0}=1$) which amounts to
the vanishing of the nonlinear correction to EM theory in the HE Lagrangian (%
\ref{2}). We observe that the cosmological constant $\Lambda $ arises (from (%
\ref{37}) and (\ref{38})) for $B_{0}\neq 1.$ The particular choice $B_{0}=1$%
, sets both $\Lambda =0$ and $\beta =0,$ which removes also the undesired
current $3$-form (\ref{40}). But this is precisely the limit of linear
Maxwell theory. In this particular magnetic gauge, in which $B_{0}\neq 0,$
expectedly $\Lambda ,$ the energy-momenta and $\mathbf{\tilde{J}}$ are all
powered by the magnetic field. It is observed from (41) and (45) that for $%
B_{0}<1$/$B_{0}>1$ we have $\Lambda >0$/$\Lambda <0,$ so that $B_{0}=1,$
i.e., the linear Maxwell limit acts as a turning point in the sign of $%
\Lambda .$ Our result is in agreement with the previous works \cite{PH, HOM}%
, in which the emergent $\Lambda $ was proportional to the product of
amplitudes of colliding waves. Here our amplitudes are same ($\sim B_{0}$)
so that $\Lambda \sim B_{0}^{2}.$ This amounts in the physical units to $%
B_{0}\sim 10T$ (recall, $1T\sim 10^{-27}m^{-1}$) which is an available
strength of magnetic fields in our universe.

\section{Conclusion}

It is shown that the emergence of null currents is inevitable in the problem
of colliding NED waves of HE Lagrangian to the first order terms sourced by
a magnetic field. From physics stand point this is not an unexpected result
since in the Lagrangian (\ref{2}) the scale invariance, i.e., the symmetry
under $x^{\mu }\rightarrow \lambda x^{\mu },$ $A_{\mu }\rightarrow \frac{1}{%
\lambda }A_{\mu },$ (for $\lambda =$constant) is broken, and as a result
null currents arise. We may comment that the emergent currents at the
boundaries ($u=v=0$) is the price paid against production of the
cosmological constant \cite{NB,PH,HOM}. Let us add that since our incoming
waves are both linearly polarized in the direction of $x$ and the resulting
wave after collision has the same polarization we don't expect the
phenomenon of birefringence. To see birefringence one must change the
polarization directions of the incoming waves to be different, which may be
the subject of a separate study. The emergence of both $\mathbf{\tilde{J}}$
on the null boundaries and $\Lambda $ for ($u>0,v>0$) may be interpreted as
a redistribution of the overall energy, or distortion of the vacuum \cite%
{PH, HOM}. Finally it is needless to state that since our spacetime is still
BS, the null singularities of the Weyl tensor remain intact. For more
details in this regard one may consult \cite{JBG}. To the future, however,
for $u>0$ and $v>0$ the spacetime is free of singularities. It can be
anticipated that similar results obtained herein follow also in different
NED models other than the one considered in this paper.

\end{document}